\documentclass[12pt]{article}

\def\fun#1#2{\lower3.6pt\vbox{\baselineskip0pt\lineskip.9pt
\ialign{$\mathsurround=0pt#1\hfil##\hfil$\crcr#2\crcr\sim\crcr}}}

\begin{document}

\date{}

%\begin{titlepage}

\title{A cosmological bound on $e^+ e^-$ mass difference}
\author{A.D. Dolgov$^{a,b,c,d}$, V.A. Novikov$^{a,b,e,f}$}

\maketitle

\begin{center}

$^a$ Novosibirsk State University, Novosibirsk, 630090, Russia\\
$^b$ Institute of Theoretical and Experimental Physics, Moscow, 113259, Russia \\
$^{c}$Dipartimento di Fisica, Universit\`a degli Studi di Ferrara, I-44100 Ferrara, Italy \\
$^{d}$Istituto Nazionale di Fisica Nucleare, Sezione di Ferrara,
I-44100 Ferrara, Italy \\
$^{e}$ National Research Nuclear University MEPhI \\
$^{f}$ Moscow Institute of Physics and Technology, 141700,
Dolgoprudny, Moscow Region, Russia

\end{center}

\begin{abstract}

We demonstrate that CPT-violation due to $e^+ e^-$ mass difference generates 
a non-zero photon mass. As a result the cosmological bounds on the photon mass 
lead to the bounds on $e^+ e^-$ mass difference which are at least  by 10 orders
of magnitude stronger than the direct experimental bound.
\end{abstract}

\newpage

There is  widely spread habit to parametrize $CPT$ violation by
attributing  different masses to particle and antiparticle (see
PDG \cite{1}). This tradition is traced to the theory of $(K-{\bar K})$-meson 
oscillation. For a given momenta ${\bf q}$ the theory
of oscillation is equivalent to a non-hermitian Quantum Mechanics
(QM) with two degrees of freedom. Diagonal elements of $2 \times
2$ Hamiltonian matrix represent masses for particle and
antiparticle. Their inequality breaks CPT-symmetry. Such  strategy
has no explicit loopholes and is still used for parametrization
of CPT-symmetry violation in  $D$ and $B$  meson oscillations.\\

Quantum Field Theory (QFT) deals with an infinite sum over all  momenta.
It is important that the set of plane waves with all possible
momenta for particle and antiparticle form a complete set of
orthogonal modes and an
arbitrary free field operator can be decomposed over this set.\\

A naive generalization of  CPT-conserving QFT to CPT-violating QFT
was made by Barenboim et al. (2001) \cite{2}. They represented a
complex scalar field as an infinite sum over modes  and attributed
different masses for particle and antiparticle

\begin{equation}
\phi(x) = \sum_{\rm\bf q}\left\{a({\rm\bf
q})\frac{1}{\sqrt{2E}}e^{-i(Et-{\bf qx})}  + b^+({\rm\bf q})
\frac{1}{\sqrt{2\tilde E}} e^{i({\tilde E}t-\rm\bf
qx)}\right\}\;\; , \label{1}\
\end{equation}
where($m, E$) and $({\tilde m}, {\tilde E})$ are masses and energies for
particle and
antiparticle with momentum ${\bf q}$ respectively.
Here the pairs $\left( a^+({\bf q}),\,a({\bf q}) \right)  $ and  
$  \left( b^+({\bf q}), \,  b({\bf q})\right) $
are creation and annihilation operators for particles and antiparticles respectively.
They obey the standard Bose commutator relations.
 In this formalism one can calculate the Wightman functions:
\begin{equation}
<\phi(x), \phi(y)^{+}> = D^{+}(x-y;m)  \label{2}\,,
\end{equation}
 \begin{equation}
<\phi(x)^+, \phi(y)> = D^{-}(x-y;{\tilde m})  \label{3}\,.
\end{equation}
They are given by the standard Lorentz-invariant Pauli-Jordan
functions but with different masses. Greenberg (2002) \cite{3}
noticed that such theory is  nonlocal and acausal. The commutator
of two fields is equal to the difference
$D^{+}(x-y;m)-D^{-}(x-y;{\tilde m})$ and  does not vanish for
space-like separation, unless the two masses are the same. In this
sense the theory is not a Lorentz-invariant one. Moreover the
Feynman propagator ($T$-product of fields) is explicitly
non-invariant. Indeed in momentum space it looks like
\begin{equation}
D_{F}(q)= \frac{1}{(2 E(\rm\bf q))} \frac{1}{(q_0 - E(\rm\bf
q))}-\frac{1}{(2 {\tilde E}(\rm\bf q))} \frac{1}{(q_0 +{\tilde
E}(\rm\bf q))}\ \label{4}
\end{equation}
and can be rewritten in invariant form only if $m=\tilde
m$.

These arguments  support a general ``theorem'' that any local
fields theory that violates $CPT$ symmetry necessarily violates Lorentz
invariance. On the other hand, one can construct a nonlocal but Lorentz
invariant theory which breaks CPT but conserves C, so the masses of
particles and antiparticles remain equal~\cite{cdnt}.

Recently we have shown that  theories with different masses for
particle and antiparticle break some  local conservation laws. In
particular they break the electric current conservation \cite{4}. 
Here we make a next step and argue that a non-zero mass
difference between a charged particle and its antiparticle, $\Delta m \neq 0$,
generates a non-zero mass of  photon. In the theory where
photon interacts with non-conserved electric current nothing
protects photon from being massive and at the first loop one gets
the non-zero $m_\gamma$:
\begin{equation}
m_\gamma^2 = C\, \frac{\alpha}{\pi}\,\Delta m^2 \,  . \label{5}
\end{equation}

The coefficient $C$ can be calculated for any given convention
about QFT with different masses for particle and antiparticle. In
paper \cite{4} we argued that there is no reasonable model for
local QFT where $m \neq \tilde m$. Correspondingly
there are  no reliable theoretical frameworks
for  calculations of  $C$. Still even with uncertain
coefficient $C$ the relation (\ref{5}) is extremely interesting.

Indeed according to the PDG~\cite{1}:
\begin{equation}
|m_{e^+} - m_{e^-}|/m_{e^-} < 8 \cdot 10^{-9}
\label{Delta-m}
\end{equation}
or $\Delta m = |m_{e^+} - m_{e^-}| < 4\cdot 10^{-3}$ eV.

Hence from relation (\ref{5}) follows that  the mass difference $\Delta m$ generates
the photon mass of the order:
\begin{equation}
m_\gamma^2 \sim C\left(\frac{\alpha}{\pi}\right) \Delta m^2  \leq
10^{-5} C \,\,{\rm eV}^2 \,.
\label{6}
\end{equation}
For any reasonable coefficient $C$ and $\Delta m$ which is not too far from
the experimental upper bound  (\ref{Delta-m}) this value of $m_\gamma$ is
huge, much larger than the existing limits.

It is interesting to reformulate relation (\ref{6}) in the opposite way, i.e. to
say that an upper bound  on the photon mass produces a bound on the
mass difference for electron and positron. As it is follows from eq. (\ref{6}):
\begin{equation}
 \Delta m_e < 20\, m_\gamma/\sqrt{C} \,,
 \label{Delta-m-m-gamma}
 \end{equation}
where we have to substitute  for $m_\gamma$
the upper limit on the photon mass. These limits and discussion of their
validity are presented in the review~\cite{goldhaber-nieto}.

The Earth based experiments give for the Compton wave length of the photon
$\lambda_C > 8\cdot 10^{7}$ cm, i.e. $m_\gamma < 3\cdot 10^{-13} $ eV, and
respectively $\Delta m < 6 \cdot 10^{-12} $~eV, nine orders of magnitude stronger
than (\ref{Delta-m}).

From the measurement of the magnetic field
of the Jupiter it follows that the Compton wave length of photon is larger than
$5\cdot 10^{10}$ cm or $m_\gamma <  4\cdot 10^{-16} $ eV, and
respectively $\Delta m < 8 \cdot 10^{-15} $~eV.

The strongest solar system bound is obtained from the analysis of the solar wind
extended up to the Pluto orbit~\cite{ryutov}:
$\lambda_C > 2\cdot 10^{13}$ cm, i.e. $m_\gamma <  10^{-18} $ eV. This is an
"official" limit present by the Particle Data Group~\cite{1}. The corresponding bound on the
electron-positron mass difference is $\Delta m < 2 \cdot 10^{-17} $~eV, which is
almost 14 orders of magnitude stronger than the direct bound on $\Delta m$.

The strongest existing bound follows from the the observation of the large scale magnetic fields
in galaxies~\cite{chibisov}: $\lambda_C > 10^{22}$ cm and $m_\gamma < 2 \cdot 10^{-27} $ eV.
Correspondingly $\Delta m < 4 \cdot 10^{-26} $ eV, which is 23 orders of magnitude stronger
than the direct limit on the electron-positron mass difference.

The galactic bound on $m_\gamma$ is subject to an uncertainty related to the way in which
the electromagnetic $U(1)$-symmetry is broken. If this symmetry is broken in the soft
Higgs-like way, the large scale magnetic fields may not be inhibited for massive photons due
to formation of vortices where the symmetry is restored~\cite{adg} and the bound presented above
would be invalidated.  However, in the case we consider here the photon mass surely does not
originate from spontaneously broken gauge $U(1)$ symmetry and the arguments of ref.~\cite{adg}
are not applicable.

Similar bounds can be derived on the CPT-odd mass differences of any other electrically charged
particles and antiparticles.

It is instructive to present an example of actual calculations of $m_\gamma$.
We stress again that there is no one sample of a local Lorentz
invariant Field Theory with non-zero $\Delta m$. Therefore  no  calculation can be done
in  a formal self-consistent way.
Here we simply start with 'a la
Barenboim-Greenberg decomposition for an electron-positron spinor
field operator $\Psi(x)$:
\begin{equation}
\Psi(x) = \sum_{\rm\bf p}\left\{a({\rm\bf p})
\frac{u(p)e^{-ipx}}{\sqrt{2\omega(p)}} + b^+({\rm\bf
p})\frac{u(-{\rm\bf p})e^{i\tilde p
x}}{\sqrt{2\tilde\omega(p)}}\right\}, \label{7}
\end{equation}
\begin{equation}
\left\{a({\rm\bf p}), a^+({\rm\bf p}^\prime)\right\} =
\delta_{{\rm\bf p}, {\rm\bf p}^\prime} \; , \;\; {\rm etc.}
\label{8}
\end{equation}

The first term in this decomposition annihilates electron with mass $m$,
while the second term  creates  positron
with mass $\tilde m$. Creation and annihilation operators obey the
standard anti-commutation relations.

We also assume the validity of the usual local product of field operators for the electric current
$$
j_\mu(x) = \bar\Psi(x)\gamma_\mu \Psi(x) \;\; .
$$
Because of the electron-positron mass difference this current is not conserved,
$\partial_\mu j(x) \neq 0$.

The Feynman propagator (T-product of fields) is a sum
of electron part contribution from the  ``Past'' and of
positron part contribution from the ``Future''. The propagator is a covariant function only when $m=
\tilde m$.

Consider  the electron-positron pair contribution into the photon
propagator, i.e. the polarization operator. Actually one can argue
that polarization operator is still given by
the textbook formula with covariant propagators with different
masses $m_1$ and $m_2$ (we take $m_1 =m$ and $m_2 = \tilde m$):
\begin{equation}
\Pi_{\mu\nu} = (ie^2)\int\frac{d^Dp}{(2\pi)^D} {\rm
Tr}\frac{1}{\hat p - m_1} \gamma_\nu\frac{1}{\hat p - \hat q -
m_2} \gamma_\mu = \tilde g_{\mu\nu}\Pi_T(q^2) +
g_{\mu\nu}\Pi_L(q^2), \label{9}
\end{equation}
where $\tilde g_{\mu\nu} = g_{\mu\nu} - q_\mu q_\nu/q^2$. This
divergent integral has to be regularized and to this end we choose the
covariant dimensional regularization.

For the standard case $m_1 = m_2$ the current is conserved and only the
transverse part of polarization operator $\Pi_T(q^2)$ is non-zero.
Since there is no $1/q^2$ singularity in the integral for
$\Pi_{\mu\nu}$, the transverse polarization operator $\Pi_T(q^2)$
has to vanish at $q^2 = 0$, i.e. $\Pi_T(q^2) \sim q^2$ at $q^2 \to
0$.  Thus $\Pi_T$ gives no contribution into the photon mass.

For non-conserved currents a longitudinal function $\Pi_L(q^2)$
would be generated. Nonzero $\Pi_L(0)\neq 0$ corresponds to
non-zero photon mass. Thus we are interested in the calculation
of $\Pi_L(q^2)$. Consider the divergence of $\Pi_{\mu\nu}$, i.e.
$q_\mu \Pi_{\mu\nu}$:
\begin{equation}
q_\mu \Pi_{\mu\nu} = q_\nu \Pi_L(q^2) = ie^2
\int\frac{d^Dp}{(2\pi)^D} {\rm Tr}\left[\frac{1}{\hat p -
m_1}\right] \gamma_\nu \left[\frac{1}{\hat p - \hat q - m_L}
\right] \hat q \;\; . \label{10}
\end{equation}
In the standard case of equal masses  $\hat q$ is equal to
the difference of two
inverse propagators and we reproduce the standard Ward identity. In our case
$\hat q$ is a difference of inverse propagators plus mass difference:
\begin{equation}
\hat q = (\hat p - m_1) - (\hat p - \hat q - m_2) + (m_1 - m_2)
\;\; . \label{11}
\end{equation}
Substituting this formula into  (\ref{10}) we get a sum of 3 integrals.
The first one is zero due to Lorentz covariance.
\begin{equation}
\int\frac{d^D p}{(2\pi)^D} {\rm Tr}\gamma_\nu \frac{1}{\hat p -
m_1} \equiv 0 \;\; . \label{12}
\end{equation}
The second one
\begin{equation}
\int\frac{d^D p}{(2\pi)^D} {\rm Tr} \gamma_\nu \frac{1}{\hat p -
\hat q - m_2} = 0 \label{13}
\end{equation}
is zero if regularization allows to make a shift of variables. The
third integral is non-vanishing:
\begin{eqnarray}
q_\nu \Pi_L(q^2) &=& ie^2(m_1 - m_2) \int\frac{d^Dp}{(2 \pi)^D}
\frac{{\rm Tr}\left[(\hat p + m_1)\gamma_\nu(\hat p - \hat q +
m_2\right]}{(p^2 - m_1^2)[(p-q)^2 - m_2^2]}= \nonumber \\
&= &4e^2(m_1 - m_2)q_\nu \int\limits_0^1 dx \int\frac{d^D
p}{(2\pi)^D} \frac{m_2 x-m_1(1-x)}{[p^2 + \Delta^2]^2} \;\; ,
\label{14}
\end{eqnarray}
where $\Delta^2 = m_1^2(1-x)+m_2^2 x - q^2 x(1-x)$.

This integral is logarithmically divergent. At this moment we can
forget about dimensional regularization.

For $q^2 = 0$ this integral is trivial and one gets that
\begin{equation}
m_\gamma^2 = \Pi_L(0) =\frac{\alpha}{2\pi}\left[m_2 -
m_1\right]^2\left[\ln\frac{\Lambda^2}{m^2} - \frac{5}{3}\right]
\;\; , \label{15}
\end{equation}
where $\Lambda$ is a cut-off, i.e. the photon mass is
divergent and has to be renormalized.  There
is no principle that protect $m_\gamma$ from being non-zero.
Formally it can be an arbitrary number. But if loops have any physical sense
for such theories this number has to be proportional to the fine structure constant, $\alpha$,
and disappears for equal mass, i.e. we arrive to eq. (5)

Similar arguments can be applied to the emergence of the nonzero graviton mass
if the energy-momentum tensor is not conserved due to different masses of particles and
antiparticles. A dimensional estimate for the graviton mass 
originating from the
electron-positron loop with unequal masses of $e^-$ and $e^+$
is $ m_g^2 \sim \Delta m^2 \Lambda^2 / m_{Pl}^2 $, where $\Lambda$
is the ultraviolet cut-off and $m_{Pl}$ is the Planck mass. It is difficult to make  any qualitative conclusion
from this result but there is  discontinuity between the zero mass limit of the theory with massive
graviton and massless General Relativity~\cite{mg-discont}. This discontinuity is quite large and
contradicts observations. A possible solution can be the  Vainshtein mechanism~\cite{vainshtein} or
one or other suggestion in modified gravity theories.

\section*{Acknowledgements}
We thank Denis Comelli for important discussion and 
acknowledge the support by the grant of the Russian Federation government 11.G34.31.0047.

\end{document}